\documentstyle[epsfig]{aipproc}

\begin{document}

\title{GRLite and GRTensorJ: Graphical user interfaces to the computer algebra system GRTensorII}

\author{Mustapha Ishak\thanks{ishak@astro.queensu.ca}, Peter Musgrave, John Mourra,
Jonathan Stern and Kayll Lake\thanks{lake@astro.queensu.ca}}

\address{Department of Physics, Queen's University, Kingston, Ontario, Canada}

\maketitle

\begin{abstract}
GRLite and GRTensorJ are first and second generation graphical user interfaces to the computer algebra system GRTensorII. Current development centers on GRTensorJ, which provides fully customizable symbolic procedures that reduce many complex calculations to ``elementary functions". Although still in development, GRTensorJ, which is now available (free of charge) over the internet, is sufficiently advanced to be of interest to researchers in general relativity and related fields.
\end{abstract}

\section*{Introduction}

GRLite\cite{grlite} and GRTensorJ are first- and second- generation graphical user interfaces to the computer algebra system GRTensorII \cite{grtensorII}. They allow students and  researchers in the area of General Relativity and related fields to perform symbolic calculations either locally or through the Web. GRLite is a calculator-style tool for evaluating more common tensors and scalars, reducing them to elementary functions. GRTensorJ provides fully customizable symbolic procedures without recompilation. These interfaces, which are open source, are written in Java and will run on any platform with browser support for JDK1.1.

\section{Description}

Any  user can initiate a GRLite or GRTensorJ session remotely on the Web by logging onto the GRTensorII home page and following the links. All that is needed is a browser with support for JDK1.1. The user initiates a session by clicking on the Java applet that opens the GRLite or GRTensorJ graphical user interfaces. They appear as in figures 1 and 2. Behind the scenes, a computer algebra session in GRTensorII is started automatically. GRLite and GRTensorJ use MapleV\cite{maplev} as the algebraic engine. They are in fact compatible with any engine that can output an ASCII stream.

\subsection{GRLite}

GRLite is a proof of concept study. It is perhaps most appropriate for beginning students as it is restricted to classical tensor analysis. The commands are performed through a predefined menu and selection of buttons. The first step is to select a spacetime. This choice automatically displays the components of the metric tensor.
The second step is to simply click on the object to be calculated. GRLite includes the 35 pre-defined functions as shown in figure 1. After the calculation is displayed the user can apply to it eight simplification procedures from the menu. There is also some support, like buffer clearing and a help system. GRLite comes with a set of  pre-defined spacetimes but also offers a graphical sub-interface for entering new spacetimes defined for a given session. The development of GRLite is complete but it will be maintained in the interim. Our current development effort concentrates on GRTensorJ.

\textbf{Figure 1}
GRLite Graphical User Interface. Calculation of the differential invariant $R_{a b c d ; e}R^{a b c d ; e}$ for the Kerr metric. At the time of writing, this calculation executes in about one second on a contemporary PC with MapleV Release 5.1 and GRTensorII 1.76.

\subsection{GRTensorJ}

GRTensorJ provides all the functionality of GRLite and much more. It has a different architecture that allows it to be  expanded and programmed by the user. The commands are accessed through  menu and sub-menu selections. The first step is to select a spacetime in coordinates or in tetrads. Then, the user can select the object(s) to be calculated from the corresponding menus. After the result of the calculation is displayed the user can apply to it any simplification procedure supported by the engine.  The help system is built into the menu system as shown in figure 2. GRTensorJ comes with a selection of spacetimes and a set of pre-defined commands. Further, the user can define  new metrics, tetrads and  procedures through the GRTensorII definition facilities. These are entered through a sub-interface with the option of saving them on the users disk space (locally) or in temporary space when used over the internet. When a session begins, GRTensorJ reads a directory on the server named TextSheets (not to be confused with ``worksheets") and builds the menu and sub-menus for the interface from the underlying structure. All sub-directories to TextSheets will appear as primary menu bar items. The names of ASCII files contained within these sub-directories will be displayed as menu selections. These files contain a sequence of commands written in the syntax of the computer algebra engine being used\cite{maplemath}. By selecting a menu item the user sends these commands to the engine.  Items to be displayed in the interface window are distinguished simply by an asterisk in the file. In other words, creating new menu items and calculation commands is as simple as creating and editing  very simple ASCII files. Yet, the file, and resultant menu item, can be the equivalent of an entire worksheet with only the result chosen for display. 

\textbf{Figure 2}
GRTensorJ Graphical User Interface. An example of the embedded  help system is shown. By the nature of the interface, the ``help" system is more of the nature of an information system.

\section{Internal Design overview}

GRLite and GRTensorJ are written in Java and have been designed using an object-oriented approach. In addition, all communication between the user and the server are object-based. GRTensorJ has a multi-layer architecture that allows the generic functionality described above. This architecture is outlined below. 

\vspace{0.3cm}
{\centering \begin{tabular}{|c|}
\noalign{\vspace{-8pt}}
\hline 
GUI (Graphical User Interface)\\
\hline 
\hline 
User Functional Interface\\
\hline 
User ICM Handler\\
\hline 
Interchange Modules (ICMs)\\
\hline 
Server ICM Handler\\
\hline 
Server Functional Interface\\
\hline 
Algebraic Engine - Server Structure\\
\hline 
\end{tabular}\par}
\vspace{0.3cm}

\section{Future development}

A number of useful features are easily added to GRTensorJ. For example, recently automatic Latex, Fortran and C output has been added\cite{papers}.  More involved is the current development of a dynamic database of solutions to Einstein's field equations.


\begin{references}
\bibitem{grlite} Lake,K 1997 ``Algebraic Computing" in {\it ``Gravitation and Relativity:At the turn of the Millennium" Proceedings of GR-15}, IUCAA, Pune, Edited by Naresh Dadhich and Jayant Narlikar pp 421-427 (gr-qc/9803072)
\bibitem{grtensorII}GRTensorII is a package that runs within MapleV. It is entirely distinct from packages distributed with MapleV and must be obtained independently. The GRTensorII software and documentation is distributed freely on the World-Wide-Web from the address {\tt http://130.15.26.63/\~{}grtensor/NewGRTensorII/}
\bibitem{maplev}MapleV is copyright Waterloo Maple Software and use of MapleV through GRLite and GRTensorJ on the Web is by special arrangement  with Waterloo Maple Software. 
\bibitem{maplemath}For example, qload(test) in GRTensorII under MapleV is equivalent to qload[test] in GRTensorM under Mathematica. At present, GRTensorJ linked to GRTensorM is not public.
\bibitem{papers}A latex call in MapleV produces output suitable for use with LaTeX 2e that appears in the applet window. We are serious when we say that you can, for example, go to the Web, do your calculation, and paste the answer into the paper you are writing!
\end{references}
\end{document}